\documentclass[twocolumn,showpacs,aps, prb]{revtex4}
\usepackage{graphicx}
\usepackage{sidecap}
\usepackage{float}
\newcommand{\bq}{\begin{equation}}
\newcommand{\eq}{\end{equation}}
\newcommand{\bn}{\begin{eqnarray}}
\newcommand{\en}{\end{eqnarray}}
\begin{document}
\title{Phonon effects on the current noise spectra and the ac conductance of a single molecular junction}
\author {Guo-Hui Ding}
\affiliation{Key Laboratory of Artificial Structures and Quantum Control (Ministry of Education), Department of Physics and Astronomy,
  Shanghai Jiao Tong University, Shanghai, 200240, China}
\author {Bing Dong}
\affiliation{Key Laboratory of Artificial Structures and Quantum Control (Ministry of Education), Department of Physics and Astronomy,
  Shanghai Jiao Tong University, Shanghai, 200240, China}

\date{\today }

\begin{abstract}
     By using nonequilibrium Green's functions and the equation of motion method, we formulate a self-consistent field theory for the electron transport through a single molecular junction (SMJ) coupled with a vibrational mode.  We show that the nonequilibrium dynamics of the phonons  in strong electron-phonon coupling regime can be taken into account appropriately in this self-consistent perturbation theory, and the self-energy of phonons is connected with the current fluctuations in the molecular junction. We calculate the finite-frequency nonsymmetrized noise spectra and the ac conductance, which reveal a wealth of inelastic electron tunneling characteristics on the absorption and emission properties of this SMJ.  In the presence of a finite bias voltage and the electron tunneling current, the vibration mode of the molecular junction is heated and driven to an unequilibrated state. The influences of unequilibrated phonons on the current and the noise spectra are investigated.

\end{abstract}
 \maketitle

\newpage

\section{introduction}

 The study of electron transport through single-molecule junctions (SMJs) have attracted great research interest in recent years.\cite{Nitzan,Galperin}  One of the prominent features of molecular junctions is the essential role played by the vibrational degrees of freedom (phonons or vibrons). \cite{Smit,Leturcq,Park,Sapmaz} The coupling between the tunneling electron and the phonons in SMJs gives rise to a variety of interesting phenomena, e.g., a strong current suppression at low bias voltage termed Franck-Condon blockade in experiments on the suspended carbon nanotube quantum dots, \cite{Leturcq} side peaks of differential conductance due to the inelastic tunneling of electron accompanied with the emission or absorption of vibrons, \cite{Park} and the ubiquitous appearance of negative differential conductance in the measured current-voltage characteristics. \cite{Sapmaz}

The theory of electron tunneling through a junction with vibrational modes was pioneered by Glazman {\it et al.}\cite{Glazman} and Wingreen {\it et al.}'s works, \cite{Wingreen} in which analytical results for transmission probability are obtained based on single particle approximation. In the last decade, a number of different methods and approximations are utilized to address this problem. \cite{Flensberg,Zhu,Mitra,ZChen,Avriller}  In the high temperature case,  the sequential tunneling of electron through the SMJ and also the cotunneling effects  were investigated within the rate equation approach, \cite{Mitra,Koch} and giant Fano factors due to the avalanche-like transport of electrons was predicted in nonlinear transport regime. \cite{Koch} In the low temperature  and the weak electron-phonon interaction (EPI) case, interesting phonon effects on the current-voltage and the shot noise characteristics of this system were predicted within perturbation theory using the nonequilibrium Green's function (GF) method.  \cite{Egger,Schmidt,Haupt,Entin-Wohlman}
The study of the strong coupling regime at low temperature is a more challenge problem. Based on a polaron transformation of the Hamiltonian, Galperin {\it et al}. \cite{Galperin2006a,Galperin2006b} proposed a self-consistent perturbation theory to treat the strong coupling regime by using the equation of motion method of the nonequilibrium
GF.  H\"artle {\it et al}. \cite{Hartle} extended this method to the treatment of the systems with the several vibrational modes, and studied the nonequilibrium vibrational effects and also the quantum interference effects on the $I$-$V$ characteristics.  Recently, the full counting statistics of  currents and the zero frequency shot noise of SMJs in the strong EPI region were also addressed. \cite{Maier,Dong2013}

In addition to the dc characteristics, the ac response and finite-frequency noise properties\cite{Blanter} of  SMJ are also important for their potential applications in future quantum devices. Recent experiments have addressed the dynamical properties for semiconductor quantum dots in the high frequency regime. \cite{Gabelli,Frey}  With the rapid progresses in the field of molecular electronics, one can expect that the ac response properties in SMJ can be probed by experiments in the near future. However, the ac properties of SMJs or quantum dots coupled with vibrational mode were only addressed in a few theoretical works\cite{Ueda,Armour} in the weak coupling regime.  To a large extent, the ac properties  and transient dynamics\cite{Riwar,Tahir} of SMJ are not extensively investigated and still remains elusive.

In this paper, we will present the results of our investigation on the finite-frequency current noise spectra and ac conductance of the SMJ system within the nonequilibrium GF formalism. \cite{Haug} A self-consistent perturbation theory for calculations of the current and the current fluctuations in the strong  EPI regime is formulated. By using the equation of motion approach, we first show that the previous self-consistent field theory \cite{Galperin2006a,Galperin2006b}can be improved by noticing the fact that the self-energy of the phonons in the SMJ is determined by the current fluctuations in this system. Then, we derive an analytical expression for the nonsymmetrized noise spectra \cite{Billangeon}by applying functional derivatives to the current formula. Consequently, the ac conductance can be obtained from the nonsymmetrized noise spectra by using the out of equilibrium fluctuation-dissipation theorem.\cite{Safi}  In our numerical calculation, we find  pronounced phonon effects on the absorption and emission spectra in the source and drain side leads and also on the ac conductance. By taking into account the phonon self-energy and relaxation effect, the effects of  phonon heating \cite{Mitra,Entin-Wohlman2010,Urban} are considered when the SMJ is driven to nonequilibrium state by a large bias voltage. We show that the unequilibrated phonons lead to a general suppression of the current and also smears out some inelastic tunneling features on the $I$-$V$ curve and the nonsymmetrized noise spectra. But negative contributions to the zero frequency shot noise by inelastic tunneling processes are obvious at intermediate values of EPI strength.

The organization of the paper is as follows: In Sec. II, the model Hamiltonian and the generic current formula are given for the system in the presence of external time-dependent measuring fields.  In Sec. III, we derive the self-consistent equations for the nonequilibrium GFs of phonon and electron. In Sec IV, we give analytical expressions for the current-current correlation functions and also the ac conductance. In Sec. V, some numerical calculations on the current noise spectra and ac conductance are presented.  Section VI is devoted to the concluding remarks.

\section { model  }
We consider a molecular quantum dot which has only one energy level (with energy $\epsilon_d$) involved in the electron tunneling process, and is also coupled to a single vibrational mode (phonon) of the molecular having the frequency $\omega_0$,  with the EPI strength $g_{ep}$.   The electron can tunnel between the molecular quantum dot and the left and the right electron leads.  Hence, this system will be described by the Anderson-Holstein model:
\bn
H&=&\sum_{k\eta}\epsilon_{k\eta}c^\dagger_{k\eta}c_{k\eta}+
 \epsilon_d d^\dagger d +\omega_0 a^\dagger a + g_{ep} d^\dagger d(a^\dagger+a) \nonumber\\
&&+\sum_{k\eta}\left [ \gamma_\eta e^{i \lambda_\eta
(t)}c^\dagger_{k\eta}d_\sigma+{\rm H.c.}\right ]\;,
\en
where $\eta=L,R$ denotes the left and right leads, and $\lambda_\eta(t)$ is an external gauge potential (measuring field)  coupled to the tunnel current from the lead $\eta$ to the molecular junction. We will show in the context that  the measuring field $\lambda_\eta(t)$ provides a convenient tool \cite{Gogolin,Ding2013} for calculating various correlation functions of currents within the Schwinger-Keldysh formulation.    It should be noted that we neglect the on-site Coulomb interaction in the molecular QD and consider a spinless electron model.  Applying a canonical transformation
\bq
\tilde H= e^S H e^{-S}\;,
\eq
with $S=g d^\dagger d(a^\dagger-a)$ and the dimensionless parameter $g=g_{ep}/\omega_0$.  The transformed Hamiltonian becomes
\bn
\tilde H&=&\sum_{k\eta}\epsilon_{k\eta}c^\dagger_{k\eta}c_{k\eta}+
 \tilde\epsilon_d d^\dagger d +\omega_0 a^\dagger a  \nonumber\\
&+&\sum_{k\eta}\left [ \gamma_\eta e^{i \lambda_\eta
(t)}c^\dagger_{k\eta}d X+\gamma^*_\eta e^{-i \lambda_\eta(t)}X^\dagger d^\dagger c_{k\eta} \right ]\;,
\en
with the phonon shift operator $X=e^{g(a-a^\dagger)}$,  $X^\dagger=e^{-g(a-a^\dagger)}$  and the renormalized energy level $\tilde\epsilon_d=\epsilon_d-g_{ep}^2/\omega_0^2$. In the transformed Hamiltonian, the direct coupling between the electron and the phonon is eliminated, but the dot-lead tunneling amplitude is modified by the phonon shift operator $X$, which is responsible for the observation of the Frank-Condon steps in the current-voltage characteristics of this SMJ.

The electric current from the lead $\eta$ to the molecular dot $I_\eta(t)=-e\langle {\frac {d N_\eta} {dt}}\rangle$,  is given by
\bq
I_\eta(t)={\frac {i e}{\hbar} }\sum_{k}\left [ \gamma_\eta e^{i \lambda_\eta
(t)}\langle c^\dagger_{k\eta}d X\rangle-\gamma^*_\eta e^{-i \lambda_\eta(t)}\langle X^\dagger d^\dagger c_{k\eta}\rangle
 \right ]\;.
\eq
It is convenient to introduce  the composite operators of electron: $\tilde d\equiv d X$ and $\tilde d^\dagger\equiv  X^\dagger d^\dagger$, and the corresponding contour-ordered GF: $\tilde G_d(t,t')=-i\langle T_c \tilde d(t) \tilde d^\dagger(t')\rangle$.
Then by equation of motion method, one can express the current as a integration on the closed-time contour over the combination of the GFs of the composite operator  and the lead's operators as follows
\bn
I_\eta(t)&=&{\frac { e}{\hbar} }\sum_{k}|\gamma_\eta|^2\int dt_1\left [e^{i \phi_\eta(t,t_1)} \tilde G_d(t,t_1)
g_{k\eta}(t_1,t)\right. \nonumber\\
&&\left.  - e^{-i \phi_\eta(t,t_1)}g_{k\eta}(t,t_1)\tilde G_d(t_1,t)\right ]\;,
\en
with the phase factor $\phi_\eta(t,t_1)=\lambda_\eta(t)-\lambda_\eta(t_1)$, and
\bq
\tilde G_d(t,t')=-i\langle T_c  d(t)X(t)X^\dagger(t')d^\dagger(t')\rangle \;,
\eq
 $g_{k\eta}(t,t')$ being the bare GF of the lead.

 \section{self-consistent perturbation theory}
 In this section, a self-consistent perturbation procedure for studying the out of equilibrium phonon and electron dynamics in this SMJ is outlined, which is based on the previous self-consistent perturbation theory suggested by Galperin {\it et al.}, \cite{Galperin2006a,Galperin2006b}  but we find that the calculations on the phonon dynamics can be improved as shown in the following.

 First, we will derive a generic relation between the phonon dynamics and the current fluctuations in this system. To study the phonon dynamics, one introduce the phonon momentum operator and phonon displacement operator
  \bq
  P_a=-i(a-a^\dagger), \hspace{1cm} Q_a=a+a^\dagger\;,
  \eq
and also the contour-ordered GF
 \bq
 D_{P_a}(t,t')=-i\langle T_c P_a(t)P_a(t')\rangle\;.
 \eq

To obtain the self-consistent equation for the phonon GF $D_{P_a}$, one can first derive the equation of motion in the Heisenberg picture for the phonon momentum and displacement operators $P_a$ and $Q_a$, respectively as
\bq
i {\frac {\partial P_a(t)} {\partial t}}=-i\omega_0 Q_a(t)\;,
\eq
\bq
i {\frac {\partial Q_a(t)} {\partial t}}=i\omega_0P_a(t)+2ig\sum_\eta j_\eta(t)\;,
\eq
where the particle current operator
\bq
j_\eta(t)=\frac {i} {\hbar} \sum_k \left [ \gamma_\eta e^{i \lambda_\eta
(t)} c^\dagger_{k\eta}d X-\gamma^*_\eta e^{-i \lambda_\eta(t)} X^\dagger d^\dagger c_{k\eta}
 \right ]\;,
\eq
which differs from the charge current operator only by the electron charge constant ($I_\eta=ej_\eta$).

Then,  introduce a differential operator
\bq
{D_{P_a}^{(0){-1}}}=-{\frac {1} {2\omega_0}}\left [ \frac {\partial^2} {\partial t^2} +\omega_0^2  \right ]\;,
\eq
and apply it to the GF $D_{P_a}$ in Eq. (8). It gives rise to
\bq
{D_{P_a}^{(0){-1}}}D_{P_a}(t,t')=\delta(t,t')-ig\sum_\eta \langle T_c j_\eta(t) P_a(t')\rangle\;.
\eq

Next,  applying the operator ${D_{P_a}^{(0){-1}}}$ again to the above equation from the right (with respect to the time variable
$t'$), we can obtain the equation of motion for $D_{P_a}$ as follows
\bn
&&{D_{P_a}^{(0){-1}}}D_{P_a}(t,t'){D_{P_a}^{(0){-1}}}=\delta(t,t'){D_{P_a}^{(0){-1}}} \nonumber\\
&&\hspace*{1cm} -ig^2\sum_{\eta\eta'} S^p_{\eta\eta'}(t,t')\;,
\en
where $ S^p_{\eta\eta'}(t,t')=\langle T_c \delta j_\eta(t)\delta j_\eta(t')\rangle $, is the particle current correlation functions between the leads $\eta$ and $\eta'$. It should be noted that here we have restricted our consideration to the steady state, and used the current conservation condition for a steady state, $\sum_\eta \langle j_\eta(t)\rangle=0$.
The above differential equation can be rewritten as an integral equation
\bn
&&D_{P_a}(t,t')=D_{P_a}^{(0)}(t,t')\nonumber\\
&&+\int dt_1 dt_2 D_{P_a}^{(0)}(t,t_1)\Pi(t_1,t_2)D_{P_a}^{(0)}(t_2,t')\;,
\en
with the self-energy term given by
\bn
\Pi(t_1,t_2)&=&-ig^2\sum_{\eta\eta'} S^p_{\eta\eta'}(t,t')
\nonumber\\
&=&-ig^2\sum_{\eta\eta'}\langle T_c \delta j_\eta(t_1)\delta j_{\eta'}(t_2)\rangle\;.
\en

As in Refs. [\onlinecite{Galperin2006a,Hartle}], we will make an approximation to replace the bare GF $D_{P_a}^{(0)}(t_2,t')$ in right hand of the above integral equation by the full phonon GF $D_{P_a}(t_2,t')$, then it gives the closed form of the Dyson equation for the phonon GF.
\bn
&&D_{P_a}(t,t')=D_{P_a}^{(0)}(t,t')\nonumber\\
&&+\int dt_1 dt_2 D_{P_a}^{(0)}(t,t_1)\Pi(t_1,t_2)D_{P_a}(t_2,t')\;.
\en

The above equation indicates that the self-energy for the GF of the phonon momentum operator is generated from  the current fluctuations through the molecular junction.
If one can express $S^p_{\eta\eta'}$ in terms of the phonon GF $D_{P_a}$ and the electron GF $G_c$, then  a self-consistent procedure of
calculation can be fulfilled. We will discuss how to obtain analytical expression of $S^p_{\eta\eta'}$ in the next section. It should be noted that the self-energy term in Eq. (16) has contributions not only from current correlations in the same leads but also from  cross current correlations between the left and the right leads.

 It is interesting to observe that the retarded(or advanced) and lesser(or greater) GFs of current operators  are directly related to that
of the phonon momentum operator in this system, for instance
\bq
D_{P_a}^{r}(\omega)=D_{P_a}^{(0),r}(\omega)+ D_{P_a}^{(0),r}(\omega)\Pi^{r}(\omega)D_{P_a}^{r}(\omega)\;,
\eq

\bq
D_{P_a}^{<}(\omega)= D_{P_a}^{r}(\omega)\Pi^{<}(\omega)D_{P_a}^{a}(\omega)\;,
\eq
Therefore, in an out of equilibrium steady state, we can study the damping effect and also the occupation number of phonons by studying the emission and absorption parts of current noise spectra in leads. This reflects the coupling and the energy flow balance in the dynamics of electron and phonon degrees of freedom in this nonequilibrium system.

using the usual non-crossing approximation, one can decouple the electron and phonon dynamics
\bq
\tilde G_d(t,t')\approx G_c(t,t')K(t,t')\;,
\eq
 where
 \bq
 G_c(t,t')=-i\langle T_c d(t)d^\dagger(t')\rangle\;,
 \eq
 \bq
 K(t,t')=\langle T_c X(t)X^\dagger(t')\rangle\;.
 \eq
 This decoupling is similar to the Born-Oppenheimer adiabatic approximation in the study of electron-nuclear dynamics of
 molecular system. One can understand that the motion of electron are influenced by the time-dependent fluctuations of potential generated by the phonon mode in the molecule, then the correlation function $K$ will account for the correlations of this kind of potential at different time. Then the correlation function $K$ will be expressed in terms of the phonon GF by using the second order cumulant expansion\cite{Galperin2006a}
 \bq
 K(t,t')=\exp\{-g^2[\langle P_a^2\rangle-iD_{P_a}(t,t')]\}\;.
 \eq

 Based on the transformed Hamiltonian of Eq. (3), we apply the equation of motion method and also the decoupling approximation in Eq. (20), it is straightforward to see that the electron GF $G_c(t,t')$  satisfies the non-equilibrium Dyson equation
 \bq
 (i{\frac \partial {\partial t}}-\tilde\epsilon_d)G_c(t,t')=\delta(t,t')+\int dt_1 \Sigma_c(t,t_1)G_c(t_1,t')
 \eq
 with the self-energy
 \bq
 \Sigma_c(t,t')=\sum_{k \eta} |\gamma_\eta|^2 g_{k\eta}(t,t')K(t',t) e^{-i\phi_\eta(t,t')}\;.
 \eq
 We introduce a quantity
\bq
\Sigma^{(0)}_\eta(t,t')=\bar\Sigma^{(0)}_\eta(t,t')e^{-i\phi_\eta(t,t')}\;,
\eq
with $\bar\Sigma^{(0)}_\eta(t,t')=\sum_{k} |\gamma_\eta|^2 g_{k\eta}(t,t')$, then
the the expression for the self-energy can be written as
\bq
 \Sigma_c(t,t')=\sum_{\eta}\Sigma^{(0)}_\eta(t,t')K(t',t)\;.
\eq
This kind of notations can simplify our formal derivation of  current and  current fluctuation formula in the next section. One sees that once the correlation function $K$ is calculated, it will be straightforward to obtain the self-energy $\Sigma_c$ and the electron GF $G_c$.

\section{ Current fluctuation and the ac conductance}

 In this section,  we will derive analytical expressions for the current fluctuations and ac conductance in this molecular junction in the framework of self-consistent perturbation theory. By using  the definition of self-energy $\Sigma^{(0)}_\eta$ in Eq. (26), it is easy to observe that the current from the lead $\eta$ to the molecule given by Eq.(5) can be rewritten as
 \bq
 I_\eta(t)={\frac {e} {\hbar}}\int dt_1 \left [ \tilde G_d(t, t_1)\Sigma_\eta^{(0)}(t_1, t)-\Sigma_\eta^{(0)}(t, t_1)\tilde G_d(t_1, t)  \right ]\;.
 \eq
Using the operational rules given by Langreth for contour integration, \cite{Haug,Langreth} one can show that this formula is equivalent to the current formula obtained by Jauho {\it et al.} for electron transport through a quantum dot,\cite{Jauho} except for  the quantum dot GF $\tilde G_d$ here is for the composite operator.

The current formula Eq. (28) can be simply expressed as
\bq
 I_\eta(t)=-i{\frac e  \hbar}  \int dt_1 dt_2
\tilde G_d(t_1,t_2)\Gamma^{(0)}_\eta(t_2,t_1;t)\;,
\eq
where we introduce the bare current vertex function $\Gamma^{(0)}_\eta(t_2,t_1;t)$  given by a functional derivative of $\Sigma^{(0)}_\eta(t_2, t_1)$ with respect to $\lambda_\eta (t)$
\begin{eqnarray}
\Gamma^{(0)}_\eta(t_2,t_1;t)&=&{\frac {\delta\Sigma_\eta^{(0)}(t_2,t_1)} {\delta\lambda_\eta(t)}}
\nonumber\\
&=&i\left [\delta(t_1,t)-\delta(t_2,t)\right ]\Sigma^{(0)}_\eta(t_2,t_1)\;.
\end{eqnarray}
Using the decoupling approximation in Eq. (20), we can rewrite the current formula as
\bq
 I_\eta(t)=-i{\frac e  \hbar}  \int dt_1 dt_2
 G_c(t_1,t_2)K(t_1,t_2)\Gamma^{(0)}_\eta(t_2,t_1;t)\;,
\eq
Making comparison the above current formula with that of the non-interacting resonant tunneling model, one can find that
the current formula for resonant tunneling model is quite similar to the above equation but without the presence of the correlation function $K(t_1, t_2)$.  We may interpret the effects of the vibration mode on the electric tunneling  in this system is represented by the renormalization of the current vertex function and also the induced self-energy in the GF $G_c$.

Now, we are ready to calculate the current-current correlation functions on the closed-time contour.
We find that these correlation functions can be obtained by functional derivative of $I_\eta(t)$ with respect to $\lambda_{\eta'}(t')$
\bq
S_{\eta\eta'}(t,t')\equiv \langle T_c\delta I_\eta(t)\delta I_{\eta'}(t')\rangle
=i e\frac {\delta I_\eta(t)}
{\delta\lambda_{\eta'}(t')}\;.\\
\eq
By using the following identity: $\frac {\delta G_c(t_1,t_2)} {\delta\lambda_{\eta'}(t')}=\int dt_3 dt_4 G_c(t_1,t_3)\frac {\delta \Sigma_c(t_3, t_4)}{\delta\lambda_{\eta'}(t')} G_c(t_4,t_2) $, and making the approximation $\frac {\delta K(t_1,t_2)} {\delta \lambda_{\eta'}(t')}\approx 0$, which corresponds to neglecting the influence  of the external measuring potential $\lambda_{\eta'}$ on the phonon dynamics, it is straightforward to obtain a generic expression for the current correlation function\cite{Ding2013}
\begin{widetext}
\bn
S_{\eta\eta'}(t,t')&=&{\frac {e^2} \hbar}\delta_{\eta\eta'}\left [G_c(t,t')K(t,t')\Sigma^{(0)}_\eta(t',t)
+\Sigma^{(0)}_\eta(t,t')K(t',t)G_c(t',t)\right ]
\nonumber\\
&&+{\frac {e^2} \hbar}\int dt_1 dt_2 dt_3 dt_4
\left [G_c(t_1,t_2)\Gamma^{(0)}_{\eta'}(t_2,t_3;t')K(t_3,t_2)G_c(t_3,t_4)K(t_1,t_4)\Gamma_{\eta}^{(0)}(t_4,t_1;t)\right ]\;.
\en
The particle current correlation function $S^p_{\eta\eta'}$, which appears in the self-energy of phonon GF in the last section, is given by the relation $S^p_{\eta\eta'}=S_{\eta\eta'}/e^2$.

By using Eq. (27) and denoting the self-energy term $\Sigma_\eta(t,t')=\Sigma^{(0)}_\eta(t,t')K(t',t)$, we can write the
the current correlation function more explicitly as
\bn
S_{\eta\eta'}(t,t')&=&{\frac {e^2} \hbar}\delta_{\eta\eta'}\left [G_c(t,t')\Sigma_\eta(t',t)
+\Sigma_\eta(t,t')G_c(t',t)\right ]
\nonumber\\
&&+{\frac {e^2} \hbar}\int dt_1 dt_2 \left [-G_c(t,t_1)\Sigma_{\eta'}(t_1,t')G_c(t',t_2)\Sigma_\eta(t_2,t)
-\Sigma_\eta(t,t_1)G_c(t_1,t')\Sigma_{\eta'}(t',t_2)G_c(t_2,t)\right.\nonumber\\
&&\left.+G_c(t,t')\Sigma_{\eta'}(t',t_1)G_c(t_1,t_2)\Sigma_\eta(t_2,t)
+\Sigma_\eta(t,t_1)G_c(t_1,t_2)\Sigma_{\eta'}(t_2,t')G_c(t',t) \right ]\;.
\en

Among various current correlation functions, the correlation function for current noise is of particular interest, since the frequency-dependent noise spectrum of current contains the intrinsic dynamics information of this quantum dot system. In a steady state without external time-dependent potential, the current fluctuations can be characterized by the nonsymmetrized noise spectra $S^>_{\eta\eta'}(\omega)$ and $S^<_{\eta\eta'}(\omega)$, which are given by the Fourier transform of the correlation function of current operators:  $S^>_{\eta\eta'}(t,t')\equiv\langle \delta I_\eta(t)\delta
I_{\eta'}(t')\rangle$ and $S^<_{\eta\eta'}(t,t')\equiv\langle \delta I_{\eta'}(t')\delta I_{\eta}(t)\rangle$, respectively.  Therefore, we have
\bq
S^>_{\eta\eta'}(\omega)=\int dt e^{i\omega(t-t')}\langle \delta I_\eta(t)\delta
I_{\eta'}(t')\rangle\;,
\eq
and the symmetry relation $S^<_{\eta\eta'}(\omega)=S^>_{\eta'\eta}(-\omega)$.
It should be noticed that the positive frequency part of $S^>_{\eta\eta'}(\omega)$ contains the information about the photon emission spectrum of this system and the negative frequency part corresponds to the absorption spectrum.  In many theoretical works, a symmetrized correlation function for current noise, $\tilde S_{\eta\eta'}(t,t') \equiv S^{>}_{\eta\eta'}(t,t')+S^{<}_{\eta\eta'}(t,t')$, are investigated. It is noted that the symmetrized noise power is given by the sum of two nonsymmetrized noise spectra: $\tilde S_{\eta\eta'}(\omega)=S^{>}_{\eta\eta'}(\omega)+S^{<}_{\eta\eta'}(\omega)$.
In recent experiments based on on-chip quantum detectors, \cite{Billangeon} it was shown that the nonsymmetrized noise spectra such as the absorption and emission parts of noise are more easy to be measured, hence we will concentre our investigation on the nonsymmetrized  noise spectra of this system in this work.

The  $S^>_{\eta\eta'}(t,t')$ is obtained straightforwardly from Eq.(34) by using Langreth's analytical continuation rules.\cite{Langreth}  In the absence of external ac potential, we can transform it to the frequency space, and  express it in terms of the Green's functions of molecular quantum dot explicitly as
\bn
S^>_{\eta\eta'}(\omega)&=&{\frac {e^2} \hbar}\int {\frac {d\omega_1} {2\pi}}\bigg \{ \delta_{\eta\eta'}
\left [ G^>_c(\omega_1+\omega)\bar\Sigma^<_\eta(\omega_1)+\bar\Sigma^>_\eta(\omega_1+\omega)G^<_c(\omega_1)\right ]
- [G_c\bar\Sigma_{\eta'}]^>(\omega_1+\omega)[G_c\bar\Sigma_\eta]^<(\omega_1)\nonumber\\
& &-[\bar\Sigma_\eta G_c]^>(\omega_1+\omega)[\bar\Sigma_{\eta'}G_c]^<(\omega_1)+ G^>_c(\omega_1+\omega)[\bar\Sigma_{\eta'}G_c\bar\Sigma_\eta]^<(\omega_1)+[\bar\Sigma_\eta G_c\bar\Sigma_{\eta'}]^>(\omega_1+\omega)G_c^<(\omega_1)  \bigg \}\;,
\en
where the self-energy term $\bar\Sigma_\eta(\omega)$ is the Fourier transform of $\bar\Sigma_\eta(t,t')=\bar\Sigma_\eta^{(0)}(t,t')K(t',t)$.  It should be pointed out that here the notation for various products of GFs and self-energy terms is the same as that in Ref.\onlinecite{Galperin2006b}, but the above current noise formula is different, since the self-energy term $\bar\Sigma_\eta(t,t')$ here is dressed by the phonon correlation function $K$,  not the bare one induced solely by the hybridization between the leads and the dot.
\end{widetext}

The ac response properties of SMJ can be probed by measuring the change of the current in the lead $\eta$ with respect to an applied
ac potential in the lead $\eta'$, which can be characterized by the ac conductance response function
\bq
G_{\eta\eta'}(t,t')=-i\theta(t-t') \langle [I_\eta(t),e N_{\eta'}(t')]\;.
\eq
Differentiating with respect to the time variable $t'$ gives
\bq
\partial_{t'}G_{\eta\eta'}(t,t')=i\delta(t-t')C_{\eta\eta'} -\chi_{I_\eta I_{\eta'}}(t,t')\;,
\eq
where the current-current response function
\bq
 \chi_{I_\eta I_{\eta'}}(t,t')=-i\theta(t-t')\langle [I_\eta(t), I_{\eta'}(t')] \rangle\;,
\eq
and the  constant $C_{\eta\eta'}= e\langle [I_\eta(t), N_{\eta'}(t)]\rangle$.
In the frequency space Eq. (38) can be written as
\bq
G_{\eta\eta'}(\omega)={\frac {1} {i\omega}} \left [ C_{\eta\eta'}-\int^\infty_{-\infty} {\frac {d\omega_1} {2\pi}}
 {\frac {[S^>_{\eta\eta'}(\omega_1)-S^<_{\eta\eta'}(\omega_1)]}{\omega-\omega_1+i 0^+} }\right ]\;.
\eq
Here the real constant $C_{\eta\eta'}$ can be calculated as
\bn
C_{\eta\eta'}&=& -i{\frac {e^2}{\hbar}}\delta_{\eta\eta'} \int \frac {d \omega_1} {2\pi} \bigg \{ \bar\Sigma^<_{\eta}(\omega_1)[G^r_c(\omega_1)+G^a_c(\omega_1)]
\nonumber\\
&& +G^<_c(\omega_1)[ \bar\Sigma^r_{\eta}(\omega_1)+ \bar\Sigma^a_{\eta}(\omega_1)]\bigg \}\;.
\en
Therefore, the real part of ac conductance is given by
\bq
Re G_{\eta\eta'}(\omega)=\frac {1} {2\omega}[S^>_{\eta\eta'}(\omega)-S^<_{\eta\eta'}(\omega)]\label{FTR}\;.
\eq
The above equation corresponds exactly to the out-of equilibrium fluctuation-dissipation theorem given in  Ref. \onlinecite{Safi}.
Hence, we can obtain the real part of ac conductance directly from the nonsymmetrized noise spectra.

\section {results and discussion}
 Now, we will carry out the numerical calculation of the current and the finite-frequency noise spectra through the SMJ by using the analytical expressions presented in the above section.  For simplicity, we consider the system with symmetric tunneling coupling to the leads, with $\Gamma_L=\Gamma_R=0.1\omega_0$, $\Gamma=(\Gamma_L+\Gamma_R)/2$, and assume the bias voltage is applied symmetrically, $\mu_{L/R}=E_F\pm eV/2$. Therefore, we can only consider the positive bias voltage case $eV \geq 0$. In the following calculation, we
 will take the phonon frequency $\omega_0=1$ as the unit of energy and the Fermi levels of the leads at equilibrium, $\mu_L=\mu_R=E_F=0$,  as the reference of energy.

 \subsection{equilibrated and undamped phonon case}
 For better understanding of the phonon effects on the electron transport in this molecular junction system, we first present the results  by assuming the phonons in equilibrium and using the bare phonon GFs.  It can describe the systems with extremely strong energy dissipation of the vibration mode to a thermal bath, e.g., a substrate or a backgate. Therefore, the oscillator restores to its equilibrium state very quickly and can be described by an equilibrium Bose
 distribution $n_B=(e^{\omega_0/T}-1)^{-1}$ at temperature $T$. The phonon shift operator GF $K(t,t')$ will be replaced by its equilibrium correlation function,\cite{Dong2013}
\bq
K(t,t')= \left (
\begin{array}{cc}
e^{-\phi(-|\tau|)} &  e^{-\phi(\tau)} \\
e^{-\phi(-\tau)} &  e^{-\phi(|\tau|)} \\
\end{array}
\right ),
\eq
where $\phi(\tau)$ is defined as ($\tau=t-t'$)
\bq
\phi(\tau) = g^2 \left [ n_B(1-e^{-i\omega_0 \tau}) + (n_B+1) (1-e^{i\omega_0 \tau}) \right ].
\eq
It is noted that in this approximation, the electron can emit or absorb phonons during the tunneling processes, but back-action of electron tunneling on the phonon distribution is neglected, except for the shift of the equilibrium position of the quantum oscillator.

\begin{figure}[htb]
\includegraphics[height=2.8in,width=\columnwidth]{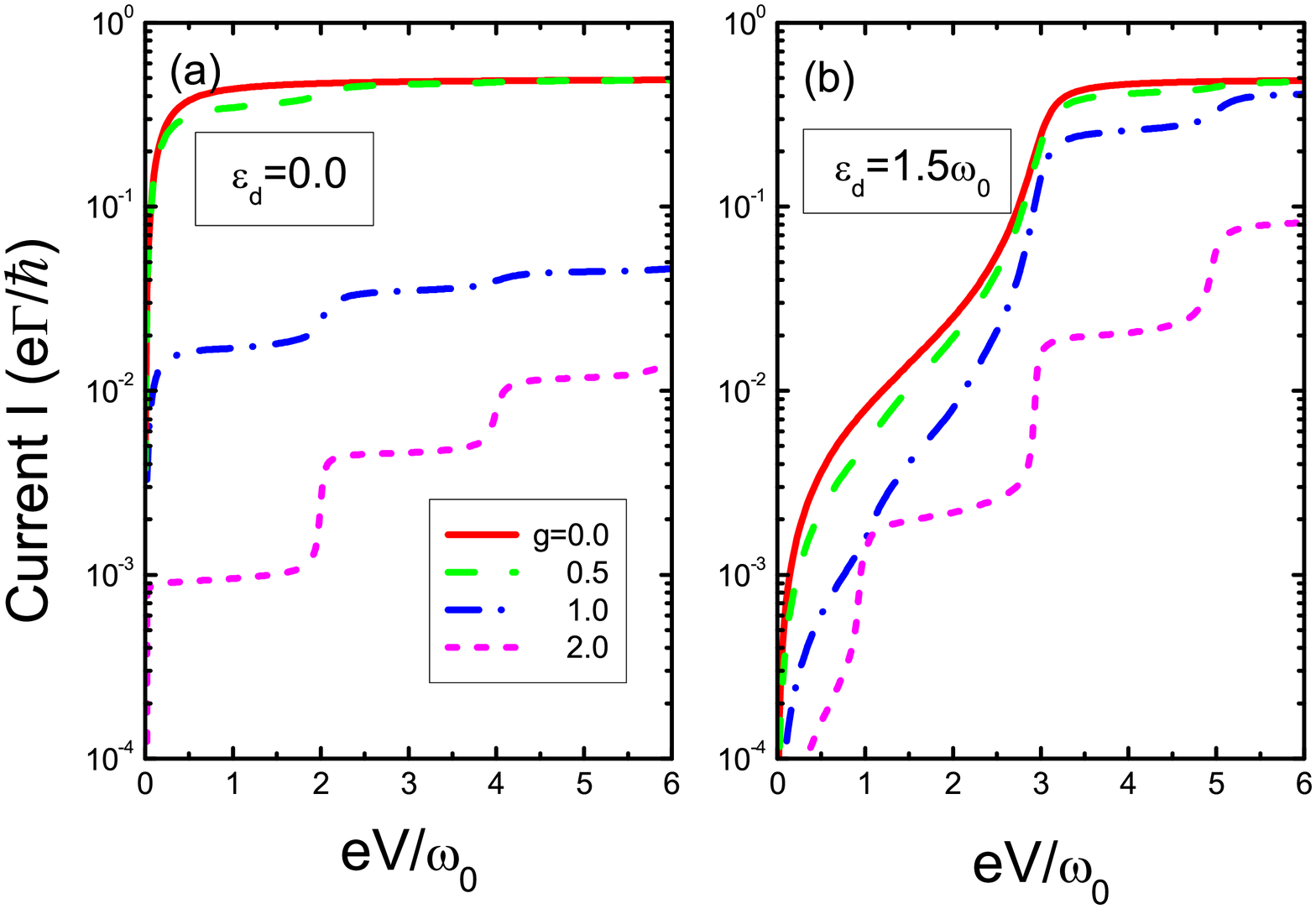}
\caption{(Colour online) The current $I$ vs the applied bias voltage $V$ at zero temperature.  The dot levels are (a) $\tilde\epsilon_d=0.0$, (b) $\tilde\epsilon_d=1.5\omega_0$, respectively. The current curves  for different values of EPI strength: $g=0.0$ (red line), $0.5$
 (green line), $1.0$ (blue line), 1.5 (cyan line), $2.0$ (magenta line), are plotted.  The other parameters used for the calculation are taken as: $\Gamma_L=\Gamma_R=0.1\omega_0$, and the chemical potential $\mu_L=-\mu_R=eV/2$.}
\label{fig1}
\end{figure}

We plot the current-voltage characteristics for different values of electron-phonon coupling strength in Fig.1. A prominent feature shown in Fig.1 (a) and (b) is the Frank-Condon blockade effect as manifested by the drastic suppression of the current with increasing the value of electron-phonon coupling constant. Therefore, for the SMJ with strong electron-phonon coupling, the electron transport is effectively blocked at low bias voltage, which was predicted by Koch {\it et al.}\cite{Koch} based on  the rate equation approach and demonstrated unambiguously in recent experiment on suspended carbon nanotube quantum dots. \cite{Leturcq} Another interesting feature is the plateau structure of the $I$-$V$ curves in the nonlinear transport regime, which is attributed to the inelastic tunneling current: when the bias voltage is greater than the phonon energy, the inelastic tunneling channel with excitations of vibrational modes is opening, which leads to the upward steps of the total current. By making comparison between Fig.1 (a) and (b), one also observes that the current is  greatly affected by the position of the energy level participated in electron tunneling processes.  For the energy level turned away from the Fermi level as shown in Fig.1 (b), the magnitude of current is significantly reduced in the low bias voltage regime.

Next, we study the current fluctuation properties in this SMJ. In the equilibrium case without bias voltage, the averaged tunneling current between the left and right leads is zero. However, the current fluctuations in the system still exist and can be detected by measuring the noise spectra. This noise spectra contain the information of the intrinsic dynamics of this system.  In Fig.2 nonsymmetrized noise spectra of the left lead is illustrated.  Since the noise power in the zero frequency limit corresponds to the well-known Johnson-Nyquist noise, $S=4k_BTG$, where $G$ is the linear conductance and $T$ is the temperature, in the zero temperature case, the noise power at zero frequency should be zero as indicated in the figure. It is also noted that for the nonsymmetrized noise spectra have nonzero values only in the positive-frequency parts, which correspond to the
absorption noise spectra, at zero temperature and in equilibrium state.   At finite frequency, pronounced
phonon effects on the noise power density are observed for the system with strong electron-phonon coupling, as indicated by the rapid increasing of noise power density at the frequency of integer multiples of the phonon frequency $\omega_0$. In the high frequency limit, it is noted that the noise power density approach a constant value of $e^2\Gamma/\hbar$ even for the systems with different coupling  constant $g$.  In the equilibrium case and for the system with symmetric tunneling couplings,  the noise spectrum in the right lead is equal to that of the left lead, therefore we only consider the noise spectra in the left lead here.


\begin{figure}[htb]
\includegraphics[height=2.8in,width=\columnwidth]{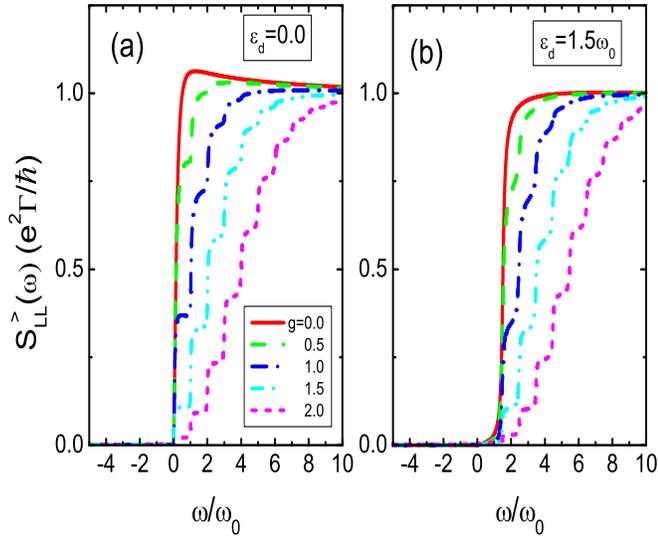}
\caption{(Colour online) The nonsymmetrized noise spectra for the left lead in the equilibrium case at zero temperature. The dot levels are (a) $\tilde\epsilon_d=0.0$, (b) $\tilde\epsilon_d=1.5\omega_0$, respectively.  The noise spectra at different values of the EPI strength $g=0.0$(red line), $0.5$(green line), $1.0$(blue line), 1.5(cyan line), $2.0$(magenta line) are plotted.  The other parameters used for calculation are taken as: $\Gamma_L=\Gamma_R=0.1\omega_0$, and the chemical potential $\mu_L=\mu_R=0.0$.}
\label{fig2}
\end{figure}

\begin{figure}[htb]
\includegraphics[height=2.8in,width=\columnwidth]{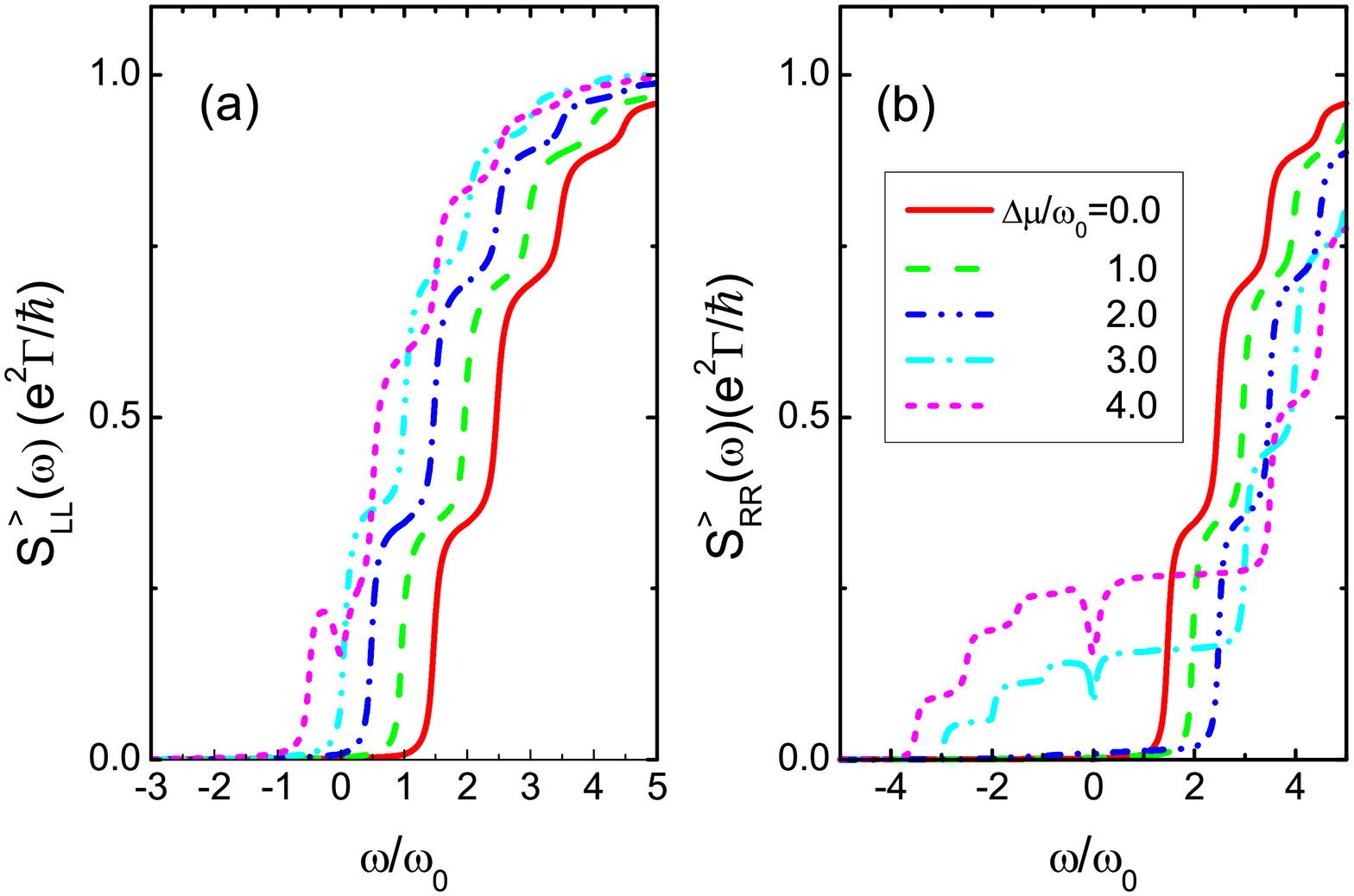}
\caption{(Colour online) The nonsymmetrized noise spectra  for the left and the right leads in the nonequilibrium case at different bias voltages. The energy level $\tilde{\epsilon}_d=1.5\omega_0$ and the electron-phonon coupling constant $g=1.0$.  The remaining parameters are the same as those in Fig.~\ref{fig1}.}
\label{fig3}
\end{figure}


For the system with finite bias voltage, the noise powers in the left and the right leads would be different. In Fig.3 (a) and (b) the nonsymmetrized noise spectra for the left and right lead at different bias voltages are plotted, respectively. Here we assume the renormalized dot level $\tilde\epsilon_d=1.5\omega_0$. In the low bias voltage region, the absorption noise power densities at positive frequency shown in Fig. 3(a) have consecutive steps at the frequencies which enable new inelastic electron tunneling channels. The jump of noise power density at consecutive steps decreases in  magnitude, which might be related to the Frank-Condon factor.  We note that with the increasing  of the bias voltage, the noise power in the left leads
shifts to the low frequency part, and it can be interpreted as a result of increasing of the chemical potential $\mu_L$ in the left lead. When the chemical potential $\mu_L$ is above the dot level $\tilde\epsilon_d$,  nonzero noise power appears in the negative frequency regime, which can be attributed to some photon emission processes during the electron tunneling from the left lead to the molecular dot. More interesting features of noise spectra are observed in the right lead side in Fig. 3(b) (the low chemical potential side). With increasing the bias voltage, the absorption noise power density merely shifts to the high frequency part at first. The reason is that in the low bias voltage regime, the bias potential is not large enough to induce significant electron tunneling through the SMJ, because the dot level $\tilde \epsilon_d$ is well above the Fermi levels.  As soon as there is bias induced electron tunneling ($\Delta\mu\ge 3\omega_0$), significant noise powers are observed in the negative frequency and the low frequency parts as shown in Fig. 3(b). We believe that the negative frequency part of noise power is resulted from the processes of energy dissipation of the tunneling electron to the drain electrode. It is interesting to notice that evident signatures of phonon effect can be found in this negative frequency part of noise spectra.  A dip structure of the noise power at zero frequency is also revealed.  Therefore, one may expect that measuring the noise power in the drain side can give rich information about the intrinsic dynamics of this system.

\begin{widetext}

\begin{SCfigure}
\includegraphics[height=3in,width=5in]{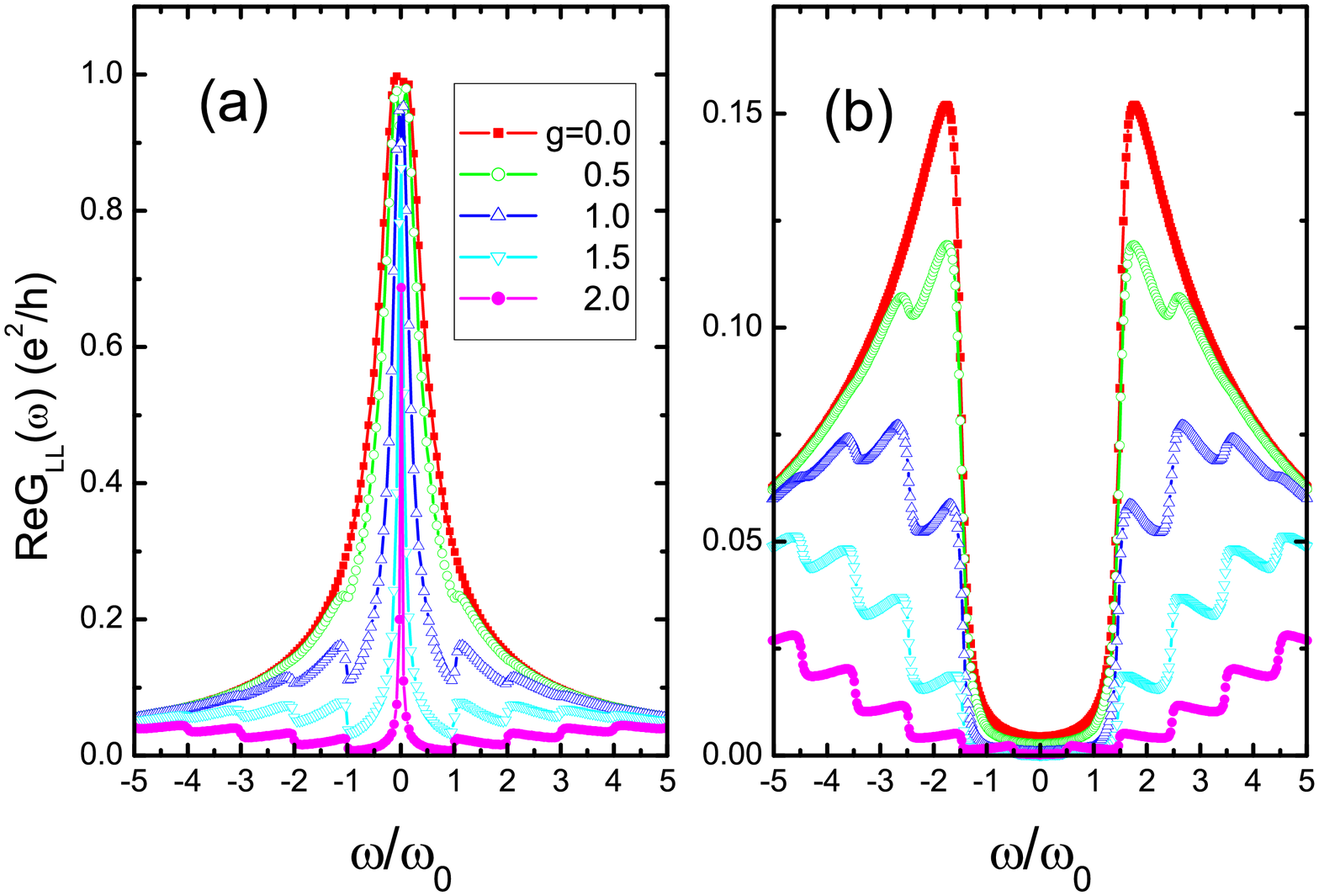}
\caption{(Colour online) The real part of the ac conductance of the left lead in the equilibrium case. (a) 
$\tilde{\epsilon}_d=0$ and (b) $\tilde{\epsilon}_d=1.5\omega_0$.  The ac conductance in the right lead is the same as
that of the left lead.}
\label{fig4}
\end{SCfigure}

\end{widetext}

\begin{figure}[htb]
\includegraphics[height=2.8in,width=\columnwidth]{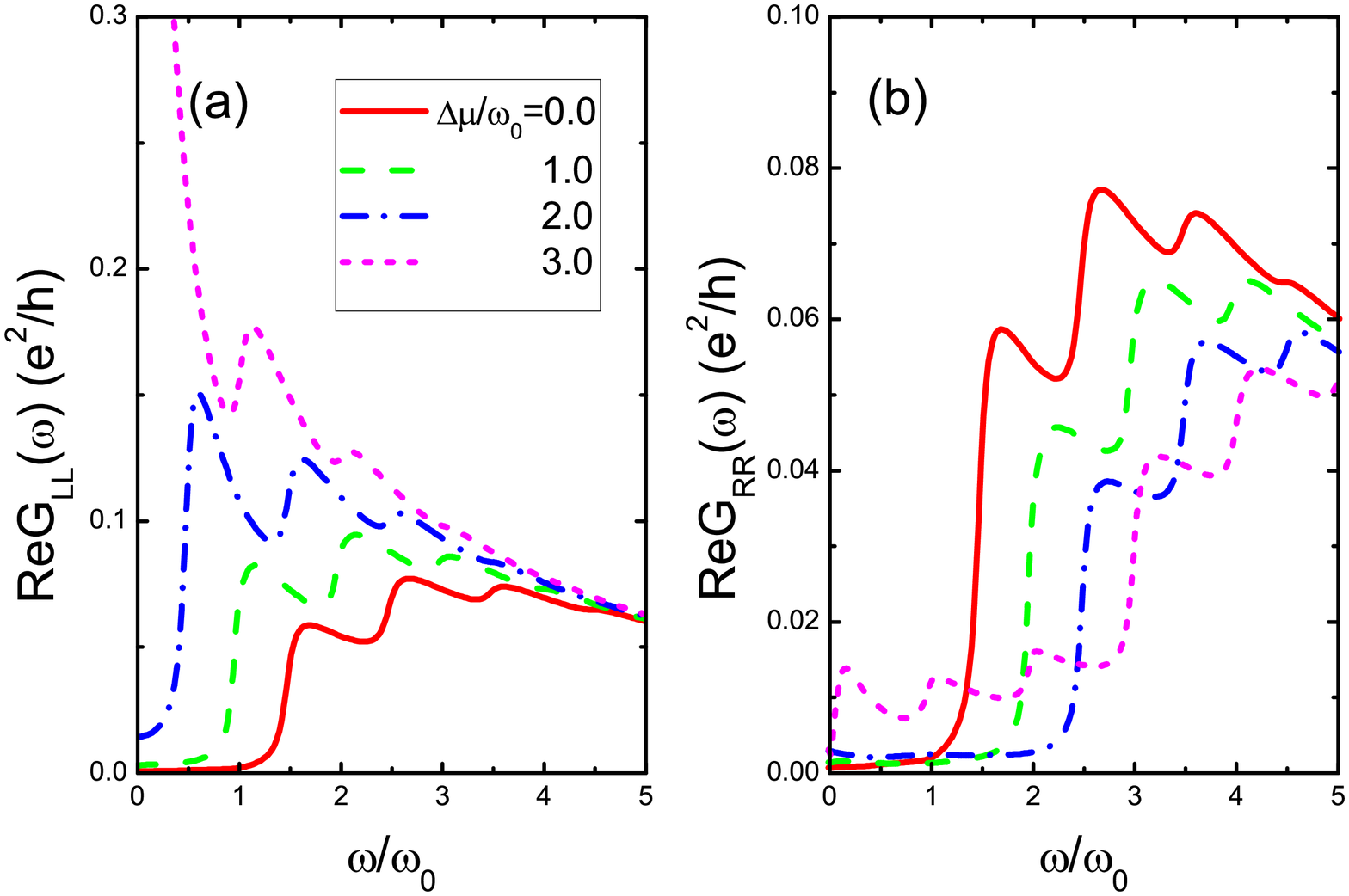}
\caption{(Colour online) The real part of ac conductance  in the out of equilibrium case at different bias voltages. (a) and (b) correspond to the left and right leads, respectively.  The renormalized dot level $\tilde \epsilon_d=1.5\omega_0$ and the EPI strength $g=1.0$. }
\label{fig5}
\end{figure}

The ac conductance is calculated from the nonsymmetrized noise spectra by using the nonequilibrium fluctuation-dissipation relation Eq. (42).
In Fig. 4 we show the ac conductance in the linear response regime of system in the equilibrium case and  at different values of EPI strength $g$. When the dot level $\tilde\epsilon_d$ is located exactly at the Fermi level of the leads,  a peak of ac conductance is found in the low frequency region as shown in Fig. 4(a), and a unitary value of conductance is achieved in the zero frequency limit. The frequency dependence of ac conductance show pronounced phonon sidebands with sudden jumps of the conductance, which can be attributed to the phonon induced logarithmic singularities in the real part of electronic self-energy and also the step structure in the imaginary part. \cite{Engelsberg,Dong2013}  The ac conductance decreases as $G_{LL}(\omega)\sim 1/\omega$ in the high frequency region. For the system with the dot level located above the Fermi level as shown in Fig. 4(b),  the maximum of the ac conductance is obtained at the frequency of finite value,  and the conductance in the low frequency regime is greatly suppressed. With increasing the coupling constant $g$, the magnitude of the ac conductance is reduced in general, whereas the phonon effect gives rise to pronounced oscillating behavior in the line shape of  the ac conductance.  Therefore, it will be an effective way to study the phonon effects in experiments by measuring the ac conductance of this SMJ systems.

For the system with finite bias voltage, the linear responses to ac potential in the left  and the right leads  are plotted in Fig.5 (a) and (b), respectively, and they show distinct features. By increasing the bias voltage, the low frequency part of the ac conductance in the left lead increases significantly, and as the Fermi level in the left lead approaches to the dot level $\tilde\epsilon_d$, a Drude peak of conductance at zero frequency is found in similar to a metallic state. On the contrary, the spectrum power of the ac conductance in the right lead is shifted to the high frequency as  the bias voltage increases.  The low frequency ac conductance is small at low bias voltage, and becomes significant when the phonons take part in the electron tunneling processes in the large bias situation. The phonon effect is manifested in the oscillation behavior of the ac conductances both in the left and right leads.

\subsection{unequilibrated phonon with self-energy }
For the system with the vibrational degrees of freedom being well isolated from the environment,  the energy dissipation of the phonons is mainly through its coupling with the leads and the tunneling electrons, therefore, the lifetime of phonons can be long enough to be driven to an unequilibrated state. In this case,  the equilibrium and bare phonon approximation might be invalid as soon as the SMJ is biased by a finite voltage and in the out of equilibrium case. The phonons in the molecular junction can be significantly heated, and they may greatly influence  the current-voltage characteristics and the noise spectra of this system.


\begin{figure}[htb]
\includegraphics[height=3.5in,width=\columnwidth]{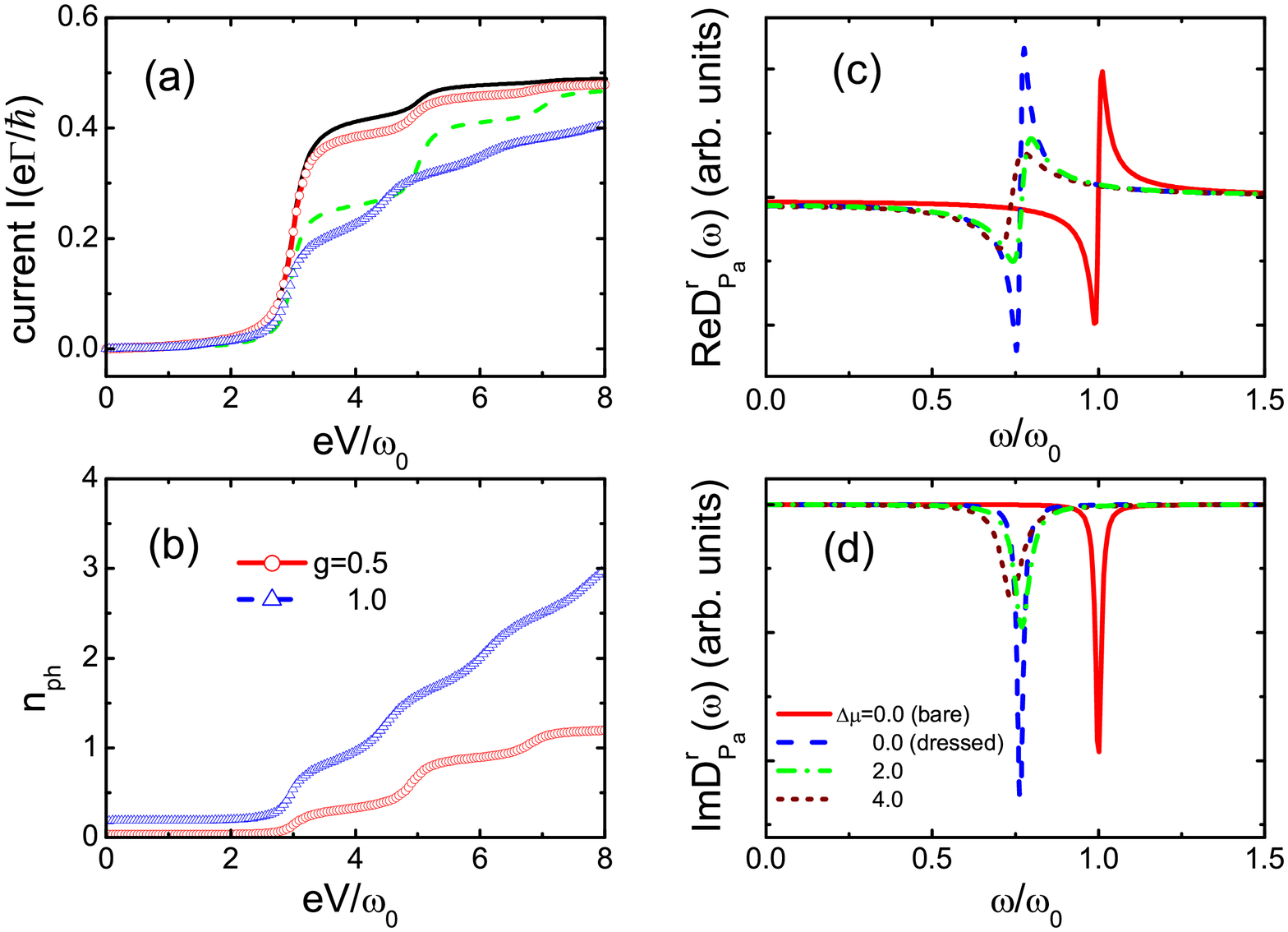}
\caption{(Colour online) (a) and (b) correspond to the current $I$ and phonon occupation number $n_{ph}$ as functions of the bias voltage, respectively, in the EPI strength $g=0.5$ and $1.0$ cases.  The black solid and green dashed line in panel (a) are the current curves in the bare phonon case.  (c) and (d) give the real and imaginary parts of the phonon GF. The dressed phonon GFs at bias voltages $\Delta\mu=0.0$ (blue dashed line), $2.0$ (green dash-dotted line), $4.0$ (wine short-dashed line) in units of $\omega_0$ are plotted and  compared with the bare phonon GF (red solid line). Here the dot level $\tilde \epsilon_d=1.5\omega_0$ and the coupling constant $g=1.0$. }
\label{fig6}
\end{figure}


By solving the self-consistent equations,  we obtain the current $I$ and phonon occupation number $n_{ph}$ vs. the bias voltage as shown in Fig. 6(a) and (b), respectively. In Fig. 6 (a) the currents obtained by taking into account the phonon damping effect  are also compared with that of the bare phonon approximation. One sees that in weak electron-phonon coupling ($g=0.5$) case, the self-consistent result for the current  only has small deviations from the bare phonon result.  With increasing the coupling constant to $g=1.0$,  significant suppression of the tunneling current by unequilibrated phonon is observed in the high bias voltage region,  and the current stepwise induced by inelastic tunneling with phonon excitations are also smeared. The occupation number $n_{ph}$ can be obtained from the lesser GF of phonon: $n_{ph}=[i\int \frac {d\omega} {2\pi} D^<_{P_a}(\omega)-1]/2$. In the bare phonon approximation and at zero temperature, $n_{ph}$ always remains zero for the system with different bias voltages.  After taking into account the phonon self-energy, the phonons on the vibration mode is driven to the out of equilibrium and  the occupation number $n_{ph}$ increases  with the bias voltage as shown in Fig. 6 (b). For the strong electron-phonon coupling case, the phonon number $n_{ph}$ increases more drastically than the weak coupling case. Therefore, the vibration mode of the SMJ is significantly heated by the applied bias between the source and drain.

In order to describe the nonequilibrium effect on vibration mode quantitatively, we plot the real and imaginary parts of the phonon GF at different bias
in Fig. 6 (c) and (d), respectively.  It is noted that the real part of phonon GF is an even function of the frequency and the imaginary part is an odd one, therefore only the positive frequency part is shown here.  We take the coupling constant $g=1.0$.  There are two distinct features for the self-consistent results of the phonon GF compared with the bare phonon case: (i) the phonon frequency is softened to the low frequency region compared with the bare phonon frequency $\omega_0$, indicating a strong renormalization effect of phonon GF; (ii) with increasing the chemical potential difference $\Delta\mu$ between the leads, the peaks for the imaginary part of GF are broadened as a result of the increased damping effect of vibrational mode in the out of equilibrium case.


\begin{figure}[htb]
\includegraphics[height=2.8in,width=\columnwidth]{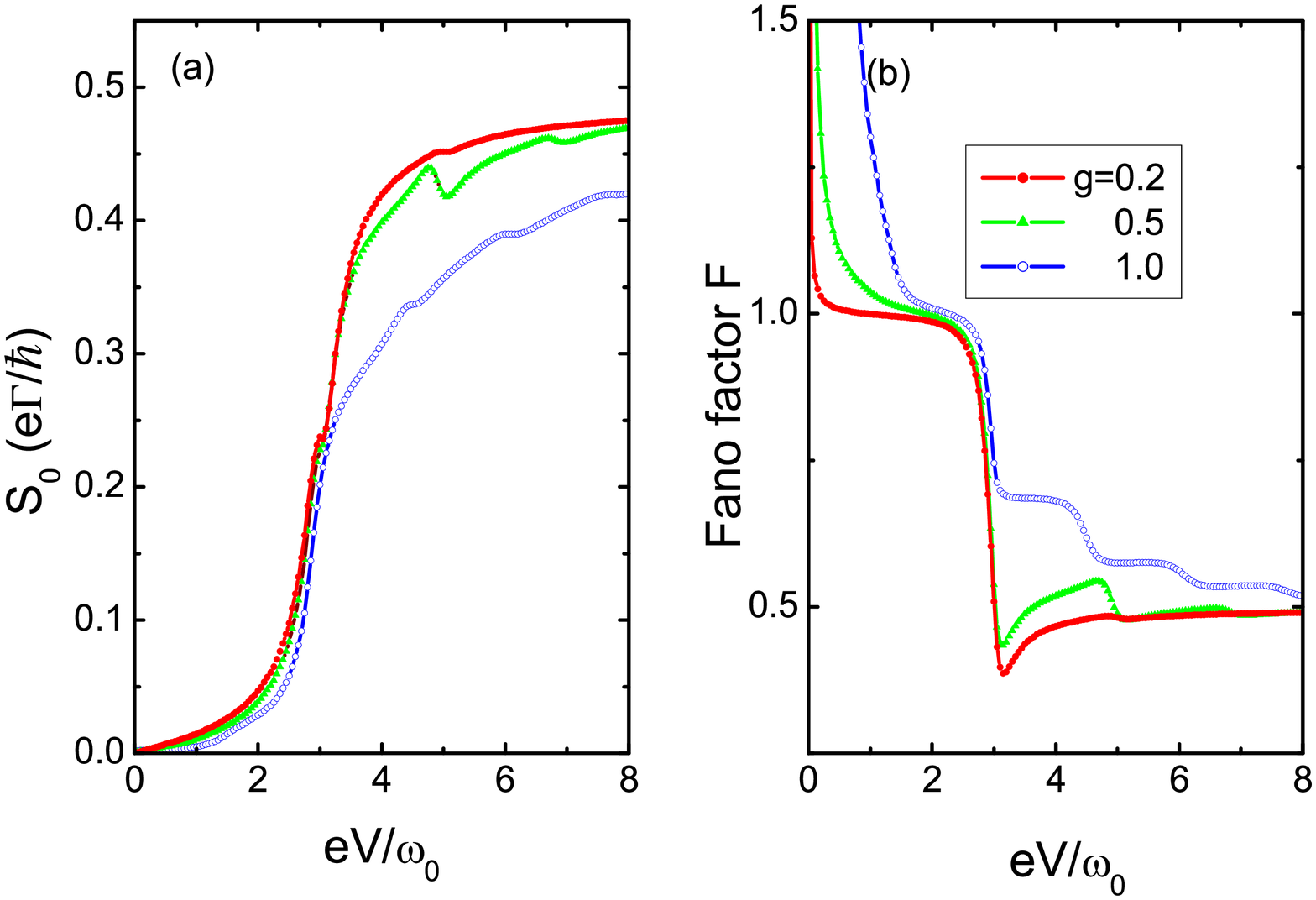}
\caption{(Colour online) The zero frequency shot noise $S_0$ and the corresponding Fano factor $F$ vs. the bias voltage $V$. Here the energy level $\tilde{\epsilon}_d=1.5\omega_0$.  }
\label{fig7}
\end{figure}


An important quantity for characterizing the fluctuations of averaged current in this SMJ is the zero frequency shot noise $S_0$, which can be calculated by the definition: $S_0=[{\tilde S}_{LL}(0)+{\tilde S}_{RR}(0)-{\tilde S}_{LR}(0)-{\tilde S}_{RL}(0)]/4$,  where ${\tilde S}_{\eta\eta'}(0)$ is the symmetrized noise power density at zero frequency. In Fig. 7(a), we plot the zero frequency shot noise $S_0$ as functions of bias voltage $V$ at different values of  coupling constants $g$. It is observed that when the chemical potential of the left lead moves above the dot level ($
eV\approx 2\tilde{\epsilon}_d=3.0\omega_0$), the shot noise increases rapidly as well as the tunneling current(shown in Fig. 6(a)).
When the chemical potential difference between the leads reaches the threshold value to excite the vibrational mode in the molecular junction,  phonon effects on the shot noise are observed in Fig. 7(a). It is interesting to find that the shot noise contributed from the inelastic tunneling processes is negative, resulting a sudden suppression of shot noise at threshold values of voltage.  This kind of negative contributions to shot noise has also been obtained in our previous calculation based on bare phonon approximation. \cite{Dong2013}  In a recent experiment on Au nanowires \cite{Kumar} with weak electron-phonon coupling,  similar negative contributions of inelastic shot noise were observed,  and they were ascribed to coherent two-electron tunneling processes mediated by phonon emission and the Pauli exclusion principles. \cite{Kumar,Schmidt} However, it should be noted that the systems consider in this paper are molecular junctions with $\Gamma\ll \omega_0$, instead of the $\Gamma\gg \omega_0$ case in the experiment on Au nanowires.  In Fig. 7(a) we find that the negative corrections of noise are most evident at intermediate values of electron-phonon interaction and will be smeared out in the strong coupling regime. It is also noted that the positions of the sudden suppressions on the shot noise are slightly shifted for the systems with different EPI strength $g$,  as a result of the renormalized phonon frequencies induced by EPI being different.    Another useful quantity to characterize the strength of noise is the so-called Fano factor $F$, which is defined as the ratio of the shot noise to the Possion value, $F=S_0/2eI$. In Fig. 7(b) the Fano factor vs. the bias voltage is plotted. The large Fano factor in the low bias voltage region is unphysical, which is due to the inaccuracy of numerical calculation, since the magnitude of current is very small in this region. Therefore, in general, the Fano factor is less than one.   One interesting feature in this electron-phonon coupling system is that the Fano factor $F$  exhibits rich oscillating structures as a function of the bias voltage. Several downward jumps of the Fano factor in conjunction with upward steps of current are observed due to the opening of new inelastic tunneling channels. In the large bias voltage limit, the Fano factors approach to 1/2 , which is in accordance with the Fano factor of a resonant tunneling model for a symmetric tunneling junction.

\begin{figure}[htb]
\includegraphics[height=2.8in,width=\columnwidth]{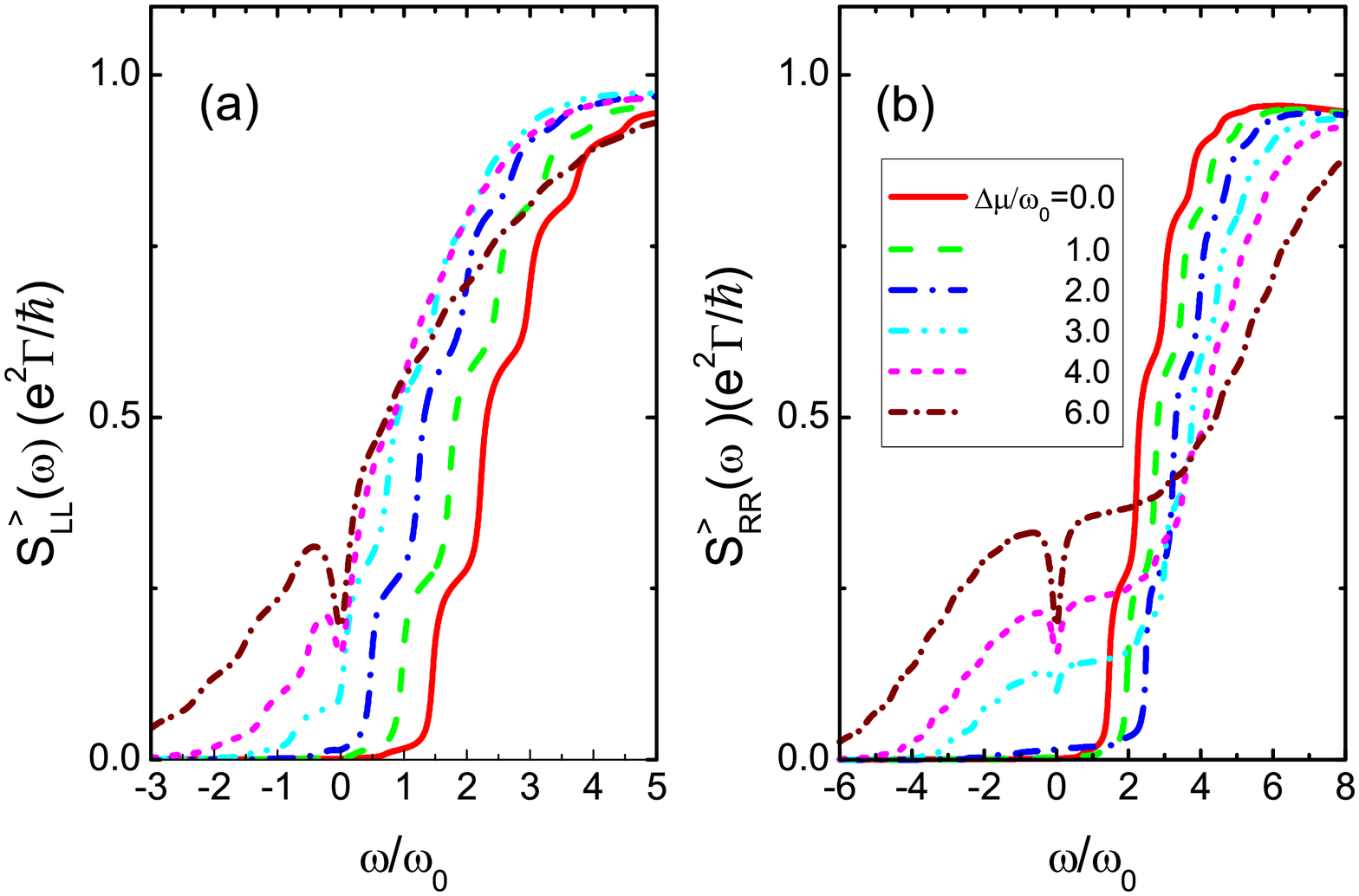}
\caption{(Colour online) The nonsymmetrized noise spectra  for the left and the right leads in the nonequilibrium case at different bias voltages. Here, the self-energy of phonon GF is taken into account by a self-consistent calculation. The energy level $\tilde{\epsilon}_d=1.5\omega_0$ and the EPI strength $g=1.0$.  }
\label{fig8}
\end{figure}

The current noise spectra for the left and right leads at different bias voltages obtained by self-consistent calculations are given in Fig. 8(a) and (b), respectively. The overall features of these nonsymmetrized noise spectra are similar to the results shown in Fig.3, where an equilibrated bare phonon
approximation is assumed.   One sees that in low bias voltage case, only absorption noise spectra in the positive frequency region are observed.  As soon as the Fermi level in the left lead  is aligned with the dot level, the tunneling current increases significantly, and the emission noise spectra in the negative frequency region becomes evident. Some signatures on the noise power density due to inelastic electron tunnelings with phonon emissions can still be observed in this calculation with the phonon self-energy  taken into account. However, the inelastic tunneling signatures on the noise spectra are not as distinct as the undamped phonon case shown in Fig.3.

\section {conclusions}
     The finite frequency current noise and the ac conductance in a SMJ have been investigated in this paper based on a self-consistent
perturbation theory. We show that the study of the self-energy and the damping effect of vibrational degree of freedom involves in the calculation of the current fluctuations and the noise spectra in the system, therefore, these physical quantities have been obtained concurrently and in a self-consistent way. The functional derivative with respect to the time-dependent counting field defined on Keldysh time-closed contour provides  a convenient tool for calculating various current-current correlation functions, which is in the same spirit as the counting field approach to the calculation of the zero frequency shot noise and also the other high order culmulants of current in charge transport through junctions. We have investigated two different situations theoretically:  At first, we assume the bare phonon GF and the equilibrium phonon approximation, It is shown that both the absorption noise spectra and the ac conductance in the linear response regime exhibit pronounced features of inelastic electron tunneling effects. In the out of equilibrium case, the emission noise spectra in the drain side of electrode show evident characteristics of phonon emissions. Secondary, the unequilibrated phonons taking into account the self-energies case is considered.  There are significantly softening of the vibration frequency of phonon due to the interaction effect and the phonon damping effects with increasing the bias voltage. The phonon occupation number becomes large in the presence of a large bias voltage, indicating the heating of vibrational modes in this molecular junction. The absorption and emission noise spectra also exhibit some features of the inelastic electron tunneling effects in this case. Recent experiments on the molecular junctions made of single-walled carbon tubes have shown strong electron-vibron coupling and high quality factor of the vibrational mode. \cite{Steele,Lassagne} We may expect the results obtained in this paper on the inelastic features of the finite-frequency noise spectra and ac conductance can be probed by future experiments on such kind of SMJs with on-chip detecting techniques.

\begin{acknowledgments}
This work was supported by Projects of the National Basic Research Program of China (973 Program) under Grant No. 2011CB925603,
the National Natural Science Foundation of China (Grant Nos.91121021 and 11074166), and Shanghai Natural Science Foundation (Grant No. 12ZR1413300).

\end{acknowledgments}


\end{document}